\documentclass[letter,prd,aps,twocolumn,floatfix,superscriptaddress,nofootinbib]{revtex4}
\usepackage{lipsum}% http://ctan.org/pkg/lipsum
\usepackage{amssymb,amsmath,graphicx}
\usepackage{hyperref}
\usepackage[usenames,dvipsnames]{color}
\usepackage{array}
\usepackage{footnote}

\makeatletter
\renewcommand*{\p@section}{\S\,}
\renewcommand*{\p@subsection}{\S\,}

\makeatother

\def\jnl@style{\it}
\def\aaref@jnl#1{{\jnl@style#1}}
\def\aaref@jnl#1{{\jnl@style#1}}

\def\aj{\aaref@jnl{AJ}}                   % Astronomical Journal
\def\araa{\aaref@jnl{ARA\&A}}             % Annual Review of Astron and Astrophys
\def\apj{\aaref@jnl{ApJ}}                 % Astrophysical Journal
\def\apjl{\aaref@jnl{ApJ}}                % Astrophysical Journal, Letters
\def\apjs{\aaref@jnl{ApJS}}               % Astrophysical Journal, Supplement
\def\ao{\aaref@jnl{Appl.~Opt.}}           % Applied Optics
\def\apss{\aaref@jnl{Ap\&SS}}             % Astrophysics and Space Science
\def\aap{\aaref@jnl{A\&A}}                % Astronomy and Astrophysics
\def\aapr{\aaref@jnl{A\&A~Rev.}}          % Astronomy and Astrophysics Reviews
\def\aaps{\aaref@jnl{A\&AS}}              % Astronomy and Astrophysics, Supplement
\def\azh{\aaref@jnl{AZh}}                 % Astronomicheskii Zhurnal
\def\baas{\aaref@jnl{BAAS}}               % Bulletin of the AAS
\def\jrasc{\aaref@jnl{JRASC}}             % Journal of the RAS of Canada
\def\memras{\aaref@jnl{MmRAS}}            % Memoirs of the RAS
\def\mnras{\aaref@jnl{MNRAS}}             % Monthly Notices of the RAS
\def\pra{\aaref@jnl{Phys.~Rev.~A}}        % Physical Review A: General Physics
\def\prb{\aaref@jnl{Phys.~Rev.~B}}        % Physical Review B: Solid State
\def\prc{\aaref@jnl{Phys.~Rev.~C}}        % Physical Review C
\def\prd{\aaref@jnl{Phys.~Rev.~D}}        % Physical Review D
\def\pre{\aaref@jnl{Phys.~Rev.~E}}        % Physical Review E
\def\prl{\aaref@jnl{Phys.~Rev.~Lett.}}    % Physical Review Letters
\def\pasp{\aaref@jnl{PASP}}               % Publications of the ASP
\def\pasj{\aaref@jnl{PASJ}}               % Publications of the ASJ
\def\qjras{\aaref@jnl{QJRAS}}             % Quarterly Journal of the RAS
\def\skytel{\aaref@jnl{S\&T}}             % Sky and Telescope
\def\solphys{\aaref@jnl{Sol.~Phys.}}      % Solar Physics
\def\sovast{\aaref@jnl{Soviet~Ast.}}      % Soviet Astronomy
\def\ssr{\aaref@jnl{Space~Sci.~Rev.}}     % Space Science Reviews
\def\zap{\aaref@jnl{ZAp}}                 % Zeitschrift fuer Astrophysik
\def\nat{\aaref@jnl{Nature}}              % Nature
\def\iaucirc{\aaref@jnl{IAU~Circ.}}       % IAU Cirulars
\def\aplett{\aaref@jnl{Astrophys.~Lett.}} % Astrophysics Letters
\def\apspr{\aaref@jnl{Astrophys.~Space~Phys.~Res.}}
\def\bain{\aaref@jnl{Bull.~Astron.~Inst.~Netherlands}} 
\def\fcp{\aaref@jnl{Fund.~Cosmic~Phys.}}  % Fundamental Cosmic Physics
\def\gca{\aaref@jnl{Geochim.~Cosmochim.~Acta}}   % Geochimica Cosmochimica Acta
\def\grl{\aaref@jnl{Geophys.~Res.~Lett.}} % Geophysics Research Letters
\def\jcp{\aaref@jnl{J.~Chem.~Phys.}}      % Journal of Chemical Physics
\def\jgr{\aaref@jnl{J.~Geophys.~Res.}}    % Journal of Geophysics Research
\def\jqsrt{\aaref@jnl{J.~Quant.~Spec.~Radiat.~Transf.}}
\def\memsai{\aaref@jnl{Mem.~Soc.~Astron.~Italiana}}
\def\nphysa{\aaref@jnl{Nucl.~Phys.~A}}   % Nuclear Physics A
\def\physrep{\aaref@jnl{Phys.~Rep.}}   % Physics Reports
\def\physscr{\aaref@jnl{Phys.~Scr}}   % Physica Scripta
\def\planss{\aaref@jnl{Planet.~Space~Sci.}}   % Planetary Space Science
\def\procspie{\aaref@jnl{Proc.~SPIE}}   % Proceedings of the SPIE

\newcommand{\sam}{{\sc MHDuet}~}
\newcommand{\samX}{{\sc MHDuet}}
\newcommand{\had}{{\sc Had}~}
\newcommand{\hadX}{{\sc Had}}
\newcommand{\lorene}{{\sc Lorene}~}

\newcommand{\samrai}{{\tt Samrai~}}
\newcommand{\samraiX}{{\tt Samrai}}

\begin{document}

\date{\today}

\title{Towards fidelity and scalability in non-vacuum mergers}

\author{Steven L. Liebling}
\affiliation{Long Island University, Brookville, New York 11548, USA}

\author{Carlos Palenzuela}
\affiliation{Departament  de  F\'{\i}sica $\&$ IAC3,  Universitat  de  les  Illes  Balears,  Palma  de Mallorca,  Baleares  E-07122,  Spain}

\author{Luis Lehner}
\affiliation{Perimeter Institute, 31 Caroline St, Waterloo, ON N2L 2Y5, Canada}

%%%%%%%%%%%%%%%%%%%%%%%%%%%%%%%%%%%%%%%%%%%%%%%%%%%%%%%%%%%%%%%%%%%%

\begin{abstract}
We study the evolution of two fiducial configurations for binary neutron stars using
two different general relativistic hydrodynamics~(GRHD), distributed adaptive mesh codes.
One code, \hadX, has for many years been used to study mergers of compact object binaries,
while a new code, \samX, has been recently developed with the experience gained with the older
one as well as several novel features for scalability improvements. As such, we examine the 
performance of each, placing particular focus on future requirements for the extraction of
gravitational wave signatures of non-vacuum binaries.
\end{abstract}

\maketitle

%%%%%%%%%%%%%%%%%%%%%%%%%%%%%%%%%%%%%%%%%%%%%%%%%%%%%%%%%%%%%%%%%%%%
\section{Introduction}\label{introduction}
%%%%%%%%%%%%%%%%%%%%%%%%%%%%%%%%%%%%%%%%%%%%%%%%%%%%%%%%%%%%%%%%%%%%

The merger of a binary neutron star~(BNS) system produces a wide range
of potentially observable signals and hence provides an ideal
event for studying high energy astrophysics through gravitational and
electromagnetic waves as well as (for sufficiently close systems) neutrinos. 
Recently, the very first merger of such a system, GW170817, was observed both by LIGO and conventional 
telescopes. Much of our understanding of this system arises from comparing
these observations with the results of numerical codes simulating such
a system. A second binary neutron star event, labeled GW190425, was detected
during O3, but only through gravitational waves due to its farther distance from the
earth. A large fraction of future BNS observations will only be detectable gravitationally,
which underscores the need for a thorough understanding of ``golden'' type events --observed
through multiple messengers so that this knowledge can be exploited in the rest of them.

Numerical simulations have been
providing an ever more detailed description of non-vacuum binaries and observational
opportunities, though the complexity of physics involved is such that, while qualitative
features are under control (prior to merger), improvements at the quantitative level are still
required.  Moreover, as the signal-to-noise ratio of detectors continues to
improve (in current and future 3rd generation detectors), increased 
accuracy will be expected from theoretically constructed
waveforms and their ability to reveal the behavior of dense matter in strongly gravitating
and dynamical regimes.

A number of codes capable of evolving a BNS through merger with fully
nonlinear general relativity exist (for reviews see Refs.~\cite{doi:10.1146/annurev-nucl-101918-023625,Radice:2020ddv}).
Such codes generally are written for distribution among a large number
of compute nodes and with the ability to refine particular spatial
regions for certain times. These implementations are called distributed adaptive
mesh refinement~(AMR) codes, and
one of such codes, \hadX, has been used extensively by the authors to model compact object
mergers (involving black holes, neutron stars and boson stars) within General
Relativity and extensions to it. A number of efforts are currently underway
to improve scalability and take advantage of massive parallel infrastructures. 
The overarching goal being to improve the study of relevant
systems exploiting advances (and related challenges) in hardware. For instance, work is already underway with discontinuous Galerkin methods~\cite{Miller:2016vik,Hebert:2018xbk} and early explorations of spectral methods
for discontinuous solutions~\cite{Piotrowska:2017pwe}. Other potential improvements have also been explored through possible GPU usage~\cite{Lewis:2018qqc}, wavelet adaptivity~\cite{Fernando:2018mov}, etc.
Clearly, there is fertile ground at several levels for considerable gains, though much effort lies ahead
to converge to the ideal set of options.

Recently, we have developed a new code
that builds on the experience gained with \had but which has a number
of important new features and improvements, called \samX. Most importantly
this code leverages the parallel infrastructure \samrai developed
at Lawrence Livermore National Laboratory to enable large scale, high resolution simulations necessary
for the GW astrophysics of non-vacuum mergers to firmly move into the quantitative regime. In this work, we compare the results for these two codes for two particular situations to assess relative accuracy of obtained solutions and discuss scalability of the new implementation.

The modeling of neutron stars requires the choice of an equation
of state~(EOS) that describes certain properties of the fluid constituting
the star and serves to close the equations of motion. 
For this comparison, we consider two particular EOSs compatible with
current constraints. In particular, we adopt the so-called
SLy EOS which is soft and  relatively compressible. The second one, MS1b,
is by contrast quite stiff.  These EOSs have been adopted by various groups, including for example
Ref.~\cite{Dudi:2018jzn} in the context of hybridization and 
Ref. ~\cite{Bernuzzi:2016pie} which discussed
errors and convergence. We note that while the MS1b  EOS is under some pressure due to EOS inferences from
gravitational wave observations~\cite{Abbott:2018wiz,Radice:2017lry,Harry:2018hke} it
nevertheless serves as a useful testbed for our purposes.

%%%%%%%%%%%%%%%%%%%%%%%%%%%%%%%%%%%%%%%%%%%%%%%%%%%%%%%%%%%%%%%%%%%%
\section{Setup}\label{details}
%%%%%%%%%%%%%%%%%%%%%%%%%%%%%%%%%%%%%%%%%%%%%%%%%%%%%%%%%%%%%%%%%%%%

We describe briefly here the evolution equations and the
numerical schemes employed by the two codes of our comparison. The Einstein equations are written as evolution equations using the 3+1 decomposition of the CCZ4 formalism~\cite{alic,bezpalen}
with constraint damping parameters $\kappa_z=0.1$ and $\kappa_c=1$. {We also use the standard 1+log and Gamma-freezing gauge conditions~\cite{bezpalen} with $\eta = 1.5$. All these damping parameters are constant in the near zone up to $r=60$ km (i.e., just beyond the region occupied by the stars) and then decay with 1/r}. The magneto-hydrodynamical~(MHD) equations describing the fluid are written as a system of conservation laws~\cite{Anderson:2007kz,Neilsen_2014}, namely
\begin{eqnarray}\label{PDEequationdecomposed}
	\partial_t {\bf U} + \partial_k {\cal F}^k({\bf U}) = {\cal S}({\bf U})
\end{eqnarray}
where ${\bf U}$ is the vector of evolved fields and ${\cal F}^k({\bf U})$ their corresponding fluxes. 
These fluxes are non-linear but they depend only on the fields themselves and not on their derivatives.
Although the full MHD system is implemented in both codes, the studies presented here sets the
initial magnetic field to zero.  

%%%%%%%%%%%%%%%%%%%%%%%%%%%%%%%%%%%
\subsection{Numerical schemes}
%%%%%%%%%%%%%%%%%%%%%%%%%%%%%%%%%%%

In both codes the discretization of the continuum equations is performed using the Method of Lines, which separates the time and space discretizations, on a regular Cartesian grid.
The spatial discretization for the evolution of the spacetime fields is performed using fourth-order accurate, spatial difference operators satisfying summation by parts~\cite{cal}, which
are based on Taylor expansions and are therefore suitable for smooth solutions. The high-frequency modes, which can not be accurately resolved by our numerical grids, are damped  by applying an artificial Kreiss-Oliger~(KO) sixth-order dissipation operator {with coefficient $\sigma=0.1$} to our evolved fields~\cite{Calabrese:2004} so as not to spoil the convergence order of our 
fourth-order numerical scheme (rigorously expected in the unigrid case for smooth solutions). 
A summary of the specific operators can be found in Ref.~\cite{Palenzuela:2018sly}.

However, these finite difference operators are not  adequate for the spatial discretization of fluxes in  genuinely non-linear systems such as GRMHD, which generically develop non-smooth solutions. For that reason, we employ High-Resolution-Shock-Capturing~(HRSC) methods~\citep{toro97,shu98}
to deal with such phenomenology in the hydrodynamical variables. 
Traditionally, most efforts in general relativistic MHD have been dedicated to the development of HRSC schemes based on finite-volumes, commonly simplified and adapted to finite-difference codes. However, in the last decade there have been several HRSC implementations based directly on finite-difference schemes, which are cheaper than finite-volume ones and can achieve high order accuracy even in multi-dimensions in an straightforward way.
In both cases (i.e, finite-volume and finite-difference approaches), the fluxes are discretized by using a conservative scheme, and the semi-discrete system in one dimension can be formally written as
\begin{eqnarray}
\label{conservative_discretization}
\partial_t {\bf U}_{i} = - \frac{1}{\Delta x} \left({\bf F}^{x}_{i+1/2} - {\bf F}^{x}_{i-1/2}\right) + {\bf S}_{i}
\nonumber 
\end{eqnarray}
where ${\bf F}^{x}_{i\pm 1/2}$ are the set of fluxes along the $x$-direction evaluated at the interfaces between two neighboring cells, located at $x_{i\pm 1/2}$. The crucial issue when dealing with shocks is how to approximately solve the Riemann problem, by reconstructing the fluxes at the interfaces such that no spurious oscillations appear in the solutions. {The differences between finite-volume and finite-difference approaches arise in the method of computation of these fluxes, and are summarized below in the description of the codes.}

An important feature of relativistic fluids is that the evolved or conserved fields are different from the physical or primitive ones appearing in the fluxes and sources. This relation is highly non-linear and the recovery of primitive fields from the conserved ones involves a numerical root solver which is one of the most delicate parts of the code. For this comparison to be meaningful, %trustful, 
we have implemented the same algorithm for the recovery, similar to the one presented in~\cite{2015PhRvD..92d4045P}. 
Another delicate issue is how to deal with the regions outside the stars, usually filled with a low density atmosphere. 
Here too, the two codes are very similar in the way in which floor values and certain energy conditions are enforced on the conserved fields.

The time integration of the resulting semi-discrete equations is performed by using either a $3^{\rm rd}$ or $4^{\rm th}$ order Runge-Kutta scheme~\cite{ShuOsherI}, which ensures the stability and convergence of the solution. We adopt a Courant parameter, defined as the ratio between the timestep and the grid size, such that $\Delta t_{l} = \lambda_c \, \Delta x_{l}$ on each refinement level $l$ to guarantee that the Courant-Friedrichs-Levy (CFL) condition is satisfied.

We next discuss details particular to each of the two codes compared here.

\subsubsection{\had}
% HAD:

The relativistic hydrodynamics equations are discretized in space by using HRSC method based on finite volumes. In particular, the fields at the interfaces are computed with a piecewise parabolic 
reconstruction~\cite{Colella:1984},
 and then
the fluxes at the interfaces are obtained from the Harten-Lax-van Leer-Einfeldt flux formula~\cite{Harten:1983,toro97}.
The time integration is performed with a third order Strong-Stability-Preserving Runge-Kutta~\cite{ShuOsherI,10.1090/S0025-5718-98-00913-2} with a CFL factor $\lambda_c = 0.25$, such that the Total Variation is preserved.

To ensure  sufficient resolution, we employ AMR via the \had computational infrastructure that provides distributed, Berger-Oliger style AMR~\cite{had,lieb} with full sub-cycling in time, together with the tapered treatment of artificial boundaries~\cite{Lehner:2005vc}. 
More details of the \had code can be found in recent studies with it of BNS mergers~\cite{2015PhRvD..92d4045P,Lehner:2016wjg,Sagunski:2017nzb}

\subsubsection{\sam}

The code \sam  has been automatically generated by using {\it Simflowny}~\cite{arbona13,arbona18}, an open-source platform to easily implement scientific dynamical models by means of a domain specific language,  to run under the \samrai infrastructure~\cite{hornung02,gunney16}, which provides parallelization and adaptive mesh refinement. We incorporated a novel AMR boundary treatment that uses an internal dense output interpolator with the information from all the RK sub-steps on the coarse grid to compute accurately the interpolated solution on the fine one~\cite{McCorquodale:2011,Mongwane:2015hja}. This algorithm (i.e., Berger-Oliger without order reduction) is not only  accurate, but also fast and efficient because uses minimal bandwidth when compared with the tapered approach of \hadX\footnote{The 
tapered approach rigorously ensures no loss of accuracy/convergence --respecting that of the
combination of time-integrator and spatial derivative operators-- for smooth solutions, though with
significant overhead and at a consequent cost in efficiency. The approach in \sam ensures 4th order accuracy
for smooth solutions.}. 
Moreover, we have extended the algorithm to allow arbitrary resolution ratios between consecutive AMR grids. The combination of this algorithm with the efficient parallelization provided by \samrai, allow us to scale at least up to 10K processes in binary black hole simulations even with high order discrete operators~\cite{Palenzuela:2018sly}. Since a GRMHD code
involves higher computational load per grid-point, scaling in the non-vacuum case is further enhanced.

We have also implemented high-order finite-difference HRSC methods to solve the fluid fields, similar to those in \cite{radice12,Bernuzzi:2016pie}. The flux formula used consist on a Lax-Friedrichs splitting~\cite{shu98}, which combines the fluxes and the fields at each node $i$ as follows:
\begin{eqnarray}\label{flux_decomposition}
F^{\pm}_{i} = \frac{1}{2} \left( F_i \pm \lambda U_i \right) 
\end{eqnarray}
where $\lambda$ is the maximum propagation speed of the system in the neighboring points. Then, we reconstruct the fluxes of each interface using the values $\{F^{\pm}\}$ from the neighboring nodes. {\sam} already incorporates some commonly used reconstructions, like the Weighted-Essentially-Non-Oscillatory (WENO) reconstructions~\cite{jiang96,shu98} and MP5~\cite{suresh97}, as well as other implementations like the FDOC families~\cite{bona09}. For the numerical simulations of binary neutron stars presented below we use MP5, which is our preferred choice for its robustness and ability to preserve sharp profiles. 

The time integration is performed with the classical fourth order  Runge-Kutta, which has only been shown to preserve the Total Variation, under certain conditions\footnote{In the way the flux operators are written.}, for values of the Courant factor below $2/3$ in 1D~\cite{ShuOsherI}. In these simulations we use $\lambda_c = 0.4$, although we also perform a simulation with $\lambda_c = 0.25$ to illustrate consistency and accuracy of the solutions obtained
within this code and also to compare with results obtained using \had.

\subsection{Initial data and EOS}

We consider the coalescence of equal mass neutron star binaries in quasi-circular orbit configuration. The initial data
is created  with \lorene using two different realistic EOSs
at zero temperature, fitted as piecewise polytropes~\cite{Read_2009}. The parameters of these EOSs, commonly known as SLy and MS1b, are summarized in Table~\ref{tab:eos}.
During the evolution we employ a hybrid EOS, where the zero  temperature effects are implemented with the piecewise polytrope, while thermal effects are modeled by an additional pressure  contribution given by the ideal gas EOS with $\Gamma= 1.75$.
The total mass  $M=2.677 M_{\odot}$ is the same for the two binaries with different EOSs, as is the initial separation $d=52.42$\,km
and the initial angular frequency $\Omega=1428$\,rad/s.

%%%%%%%%%%%%%%%%%%%%%%%%%%%%
\begin{table}[h]
	\centering
	\caption{Characterization of the two different EOSs used in this work. Each EOS is defined as a piecewise polytrope with $n=4$ segments and with $ K_0[CGS] =
		3.59389\times 10^{13}$ and $\Gamma_0 = 1.35692$.  
      Each segment is delineated by a transition density $\rho_i$
      expressed in \texttt{cgs} units.
	}
	\begin{tabular}{ccccccc}
		\hline
		EOS & $\Gamma_1$ & $\Gamma_2$ & $\Gamma_3$ & $\log_{10} \rho_0$ & $\log_{10} \rho_1$
		& $\log_{10}\rho_2$\\
		\hline
		SLy  &3.005&2.988&2.851&14.165&14.7&15.0\\
		MS1b &3.456&3.011&1.425&14.05556938&14.7&15.0\\
		\hline
	\end{tabular}
	\label{tab:eos}
\end{table}
%%%%%%%%%%%%%%%%%%%%%%%%%%%%

\subsection{Analysis}

For each of the simulations, we calculate a number of quantities as described here. We compute a retarded time $u$ in terms of the tortoise coordinate $r_*$ as
  \begin{align}
    u   &= t-r_* \\
    r_* &= R + 2 M \log\left(\frac{R}{2M}-1\right)\\
    R   &= r \left( 1 + \frac{M}{2r}\right)^2 \ , 
  \end{align}
  where $R$ is Schwarzschild radius and $r$ the isotropic radius. In
terms of our Cartesian coordinates, the isotropic radius is simply
  $r^2=x^2+y^2+z^2$.

The dominant mode of the gravitational wave strain $h_{22}$ is presented as
  \begin{equation}
     rh_{22} = A {\rm e}^{-{\rm i} \phi}
  \end{equation}
  in terms of which the gravitational wave angular frequency is defined as
  \begin{equation}
  \omega_{22} = \frac{d\phi}{dt} = - \Im{\left(\frac{\dot{h}_{22}}{h_{22}}\right)}.
  \end{equation}

%%%%%%%%%%%%%%%%%%%%%%%%%%%%%%%%%%%
\section{Results}
%%%%%%%%%%%%%%%%%%%%%%%%%%%%%%%%%%%

Let us first overview the numerical setup and the results of the numerical simulations. Each binary was evolved within
\had at two different resolutions, as shown in Table~\ref{tab:sim}.
The number of levels and most AMR parameters are identical for both runs, but the
higher resolution run adopted a resolution 50\% higher than, and a threshold
for refinement half that of, the medium resolution run.
The total number of levels of refinement for all these runs was five with the finest
resolution achieved of $\Delta x=0.25$ (in code units) $=0.38$\,km. 

As also shown in Table~\ref{tab:sim}, the \sam runs differed only in their overall resolution, with a 25\% increase between consecutive  resolutions. These simulations use five refinement levels, achieving a finest resolution $\Delta x=160 m$.

%%%%%%%%%%%%%%%%%%%%%%%%%%%%
\begin{table}[h]
	\centering
	\caption{Details of the various simulations presented here.
   For each simulation, the name of the code used, the name of
   the EOS characterizing the fluid, the name we use to refer to
   the simulation, the number of levels, the number of points in
   each direction for the coarse level, the finest grid spacing 
   achieved, and finally the CFL ratio of the timestep to grid spacing
   are displayed.} 
	\begin{tabular}{|c|c|c|c|c|c|c|}
		\hline
		Code  & EOS  & Name       & \# levels & $N$ & $\Delta x$ & $\Delta t/\Delta x$ \\ 
		\hline
		\had  & SLy  & low (L)    & 5         &  81 & 0.37 & 0.25  \\
		\had  & SLy  & medium (M) & 5         & 121 & 0.25 & 0.25  \\
		\cline{2-7}
		\had  & MS1b & low (L)    & 5         &  81 & 0.37 & 0.25  \\
		\had  & MS1b & medium (M) & 5         & 121 & 0.25 & 0.25  \\
		\hline
		\sam  & SLy  & low (L)    & 5         & 154 & 0.208 & 0.25  \\		
		\sam  & SLy  & low (L)    & 5         & 154 & 0.208 & 0.4  \\
		\sam  & SLy  & medium (M) & 5         & 192 & 0.167 & 0.4  \\
		\sam  & SLy  & high (H)   & 5         & 240 & 0.134 & 0.4  \\
		\sam  & SLy  & finest (F) & 5         & 300 & 0.106 & 0.4  \\
		\cline{2-7}
		\sam  & MS1b & low (L)    & 5         & 154 & 0.208 & 0.4  \\
		\sam  & MS1b & medium (M) & 5         & 192 & 0.167 & 0.4  \\
		\sam  & MS1b & high (H)   & 5         & 240 & 0.134 & 0.4  \\
		\hline
	\end{tabular}
	\label{tab:sim}
\end{table}
%%%%%%%%%%%%%%%%%%%%%%%%%%%%

\begin{figure*}
	\includegraphics[height=4.0in]{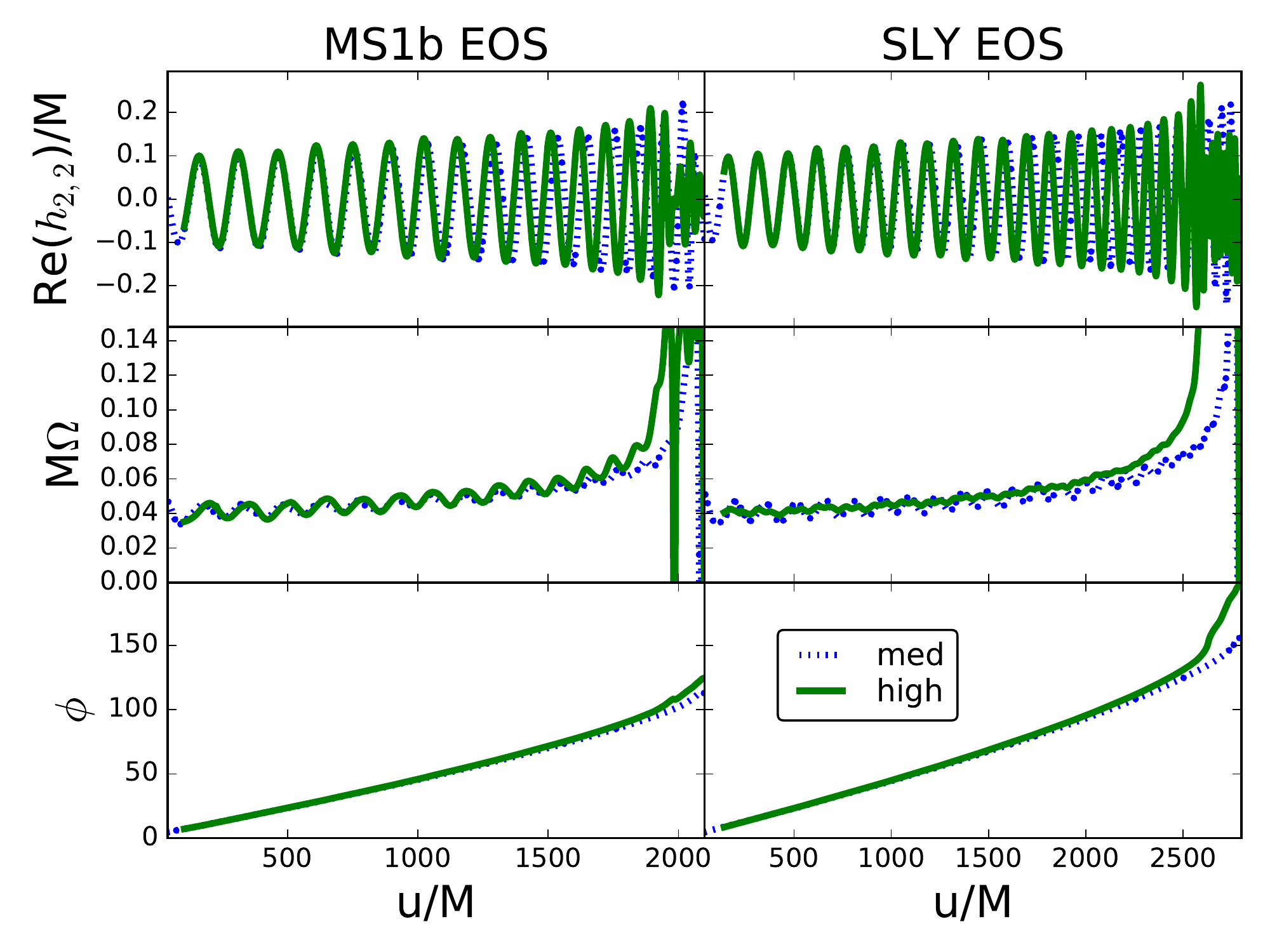}
	\caption{\had evolutions for both EOSs and different resolutions.
            The GW strain is shown in the \textbf{top} frame,
            the GW frequency is shown in the \textbf{middle}, and
            the GW phase is shown at \textbf{bottom}.
	}
	\label{fig:conv_had}
\end{figure*}

Results from the \had runs are shown in Fig.~\ref{fig:conv_had}.
Apparent in the figure, the evolutions agree at early times, but the medium resolution merges a bit earlier than the low resolution for both EOSs.

\begin{figure*}
\includegraphics[height=4.0in]{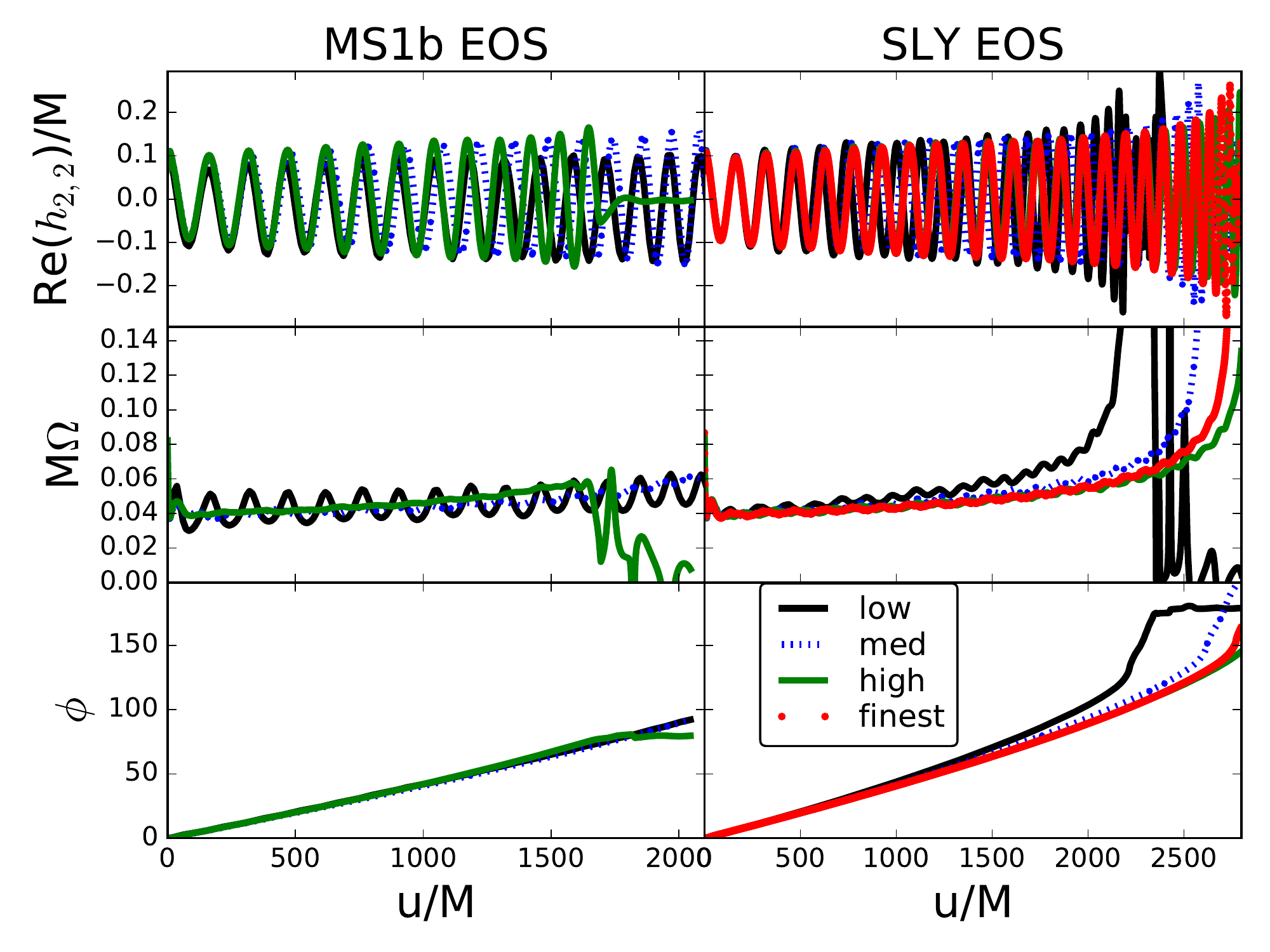}
\caption{\sam evolutions for both EOSs and different resolutions.
Between any two runs, the resolution increases by 25\%.
	Quantities shown are the same ones shown in Fig.~\ref{fig:conv_had}.
}
\label{fig:conv_sam}
\end{figure*}

The \sam runs were similar to the
\had runs in that the SLy case appears to be closer to the convergent regime than the MS1b case which 
clearly requires higher resolutions, as shown in Fig.~\ref{fig:conv_sam}.

\begin{figure*}
	\includegraphics[height=4.0in]{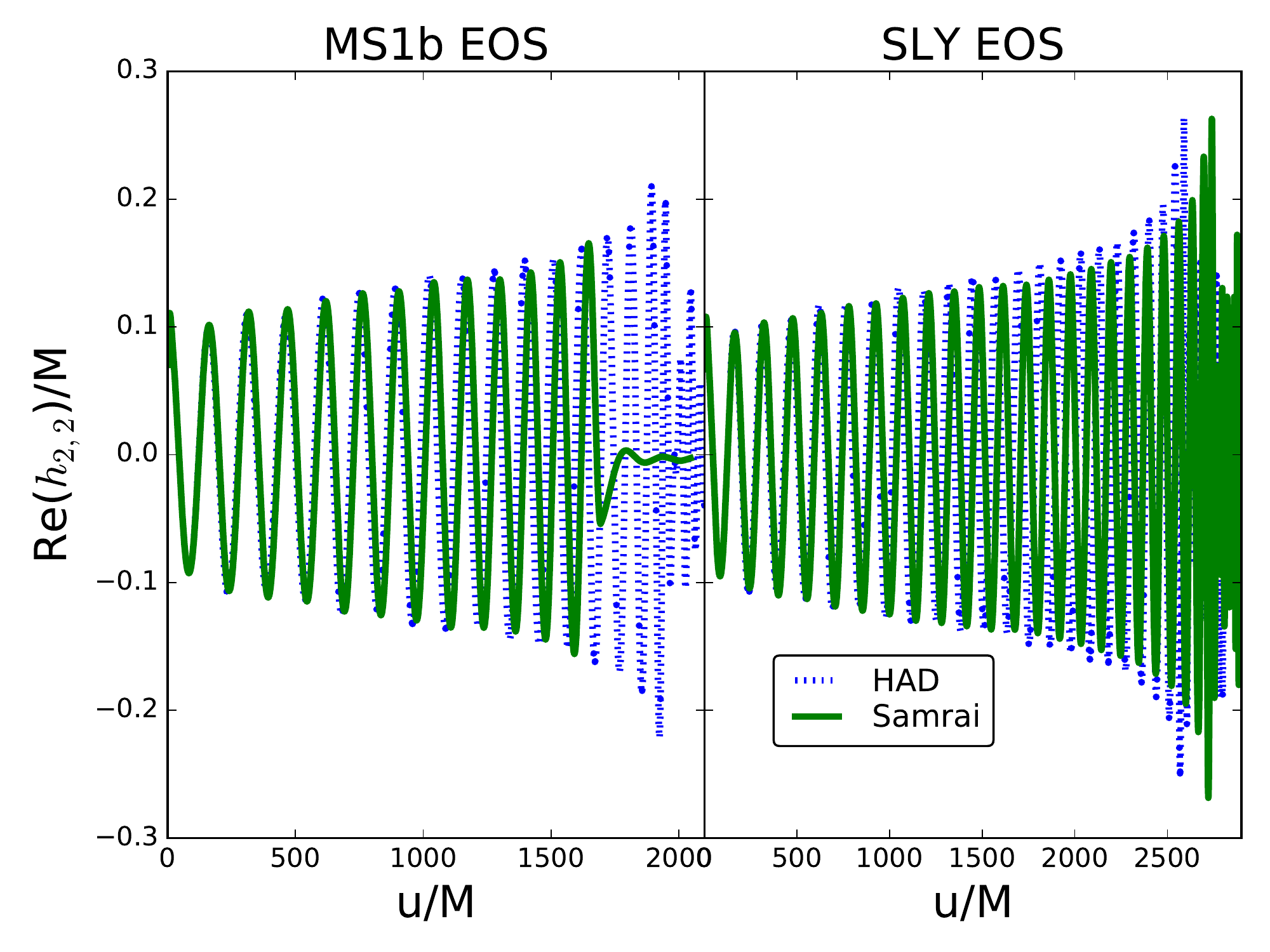}
	\caption{Strains for the highest resolution runs of the two different codes and two different EOSs.
	}
	\label{fig:strains_comparison}
\end{figure*}

A direct comparison of the highest resolution simulation available for each binary and for each code can be found in Fig.~\ref{fig:strains_comparison}.
Clearly, the deviations increase as the binary proceeds on their orbits, being more significant for the MS1b EOS. Note though that the \sam runs generally have  higher resolution than the \had runs.

\begin{figure*}
\includegraphics[height=4.0in]{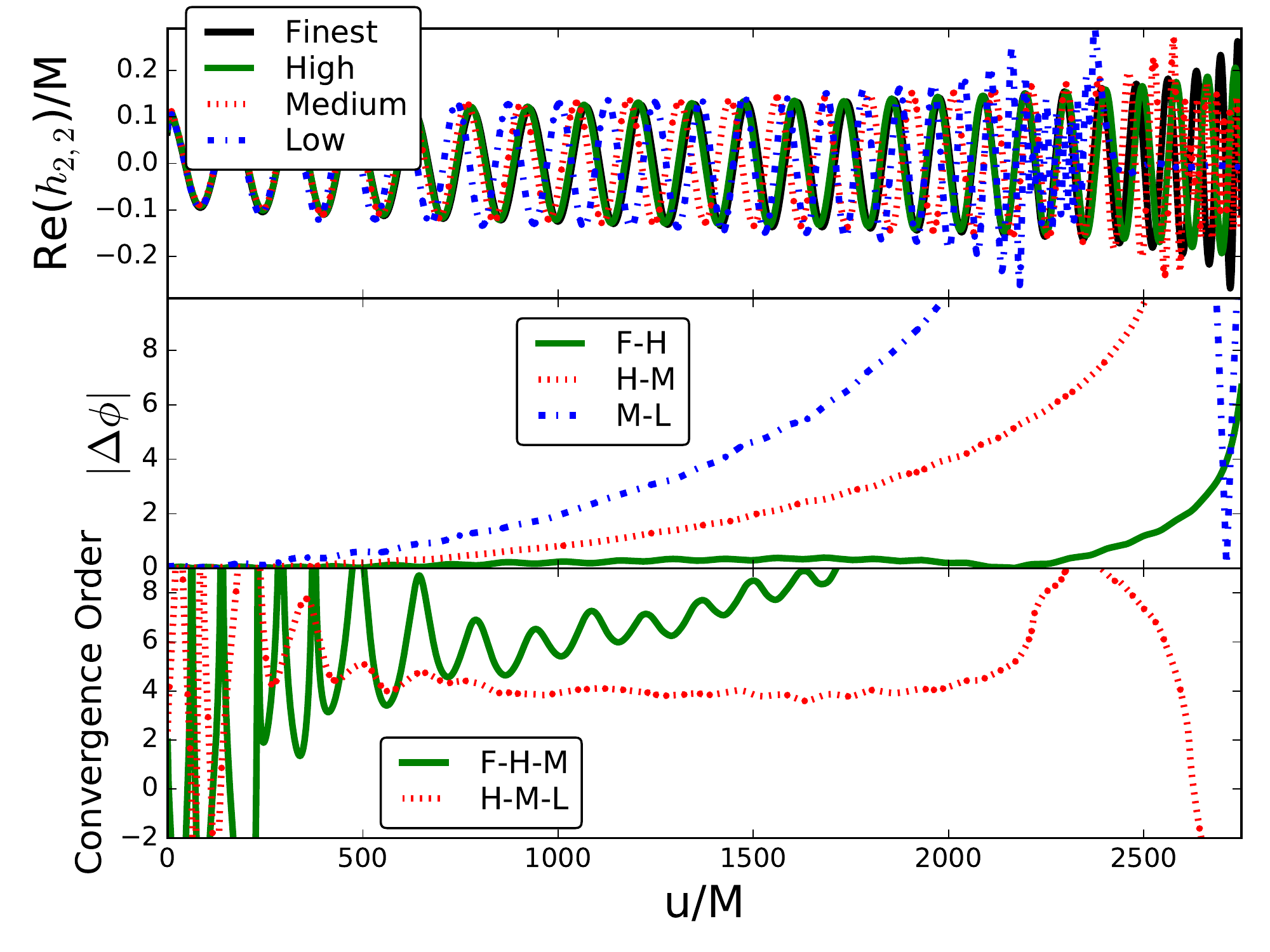}
	\caption{Errors in phase between the different resolutions for the  SLy \sam runs.
         \textbf{Top:} The strains for the four different resolutions.
         \textbf{Middle:} Differences in phase between successive resolutions.
                          These differences decrease toward zero as resolution
                          increases indicating convergence.
         \textbf{Bottom:} The convergence order estimated with just the
                 phase differences, indicating high order convergence. The early significant
                 variations are a consequence of differences being too small.
}
	\label{fig:phaseerrors}
\end{figure*}

Because we have four different resolutions for the SLy case with the \sam code,
we can study in detail how these errors behave.
In Fig.~\ref{fig:phaseerrors} we display in the top panel the waveform for the different resolutions. The middle panel 
displays the difference in phase between the different resolutions. 
Clearly, it converges to zero rapidly. The bottom panel shows that, for the lowest resolution triad, it converges roughly to fourth order, while it converges to a faster rate for the highest resolution one.
This convergence was measured only with the phase of a radiative signal,
but nevertheless such high order convergence is not typically observed in BNS
simulations (for reference see, e.g.~\cite{Dietrich:2018upm,Baiotti:2011am}). 
The high convergence order observed in the wave zone can be attributed to a couple  factors that go along
with the use of high order methods: the use of adaptive re-gridding designed to satisfy an error threshold
and the fact that the fluid discontinuities are not volume filling.
%}

\begin{figure}
\includegraphics[height=2.6in]{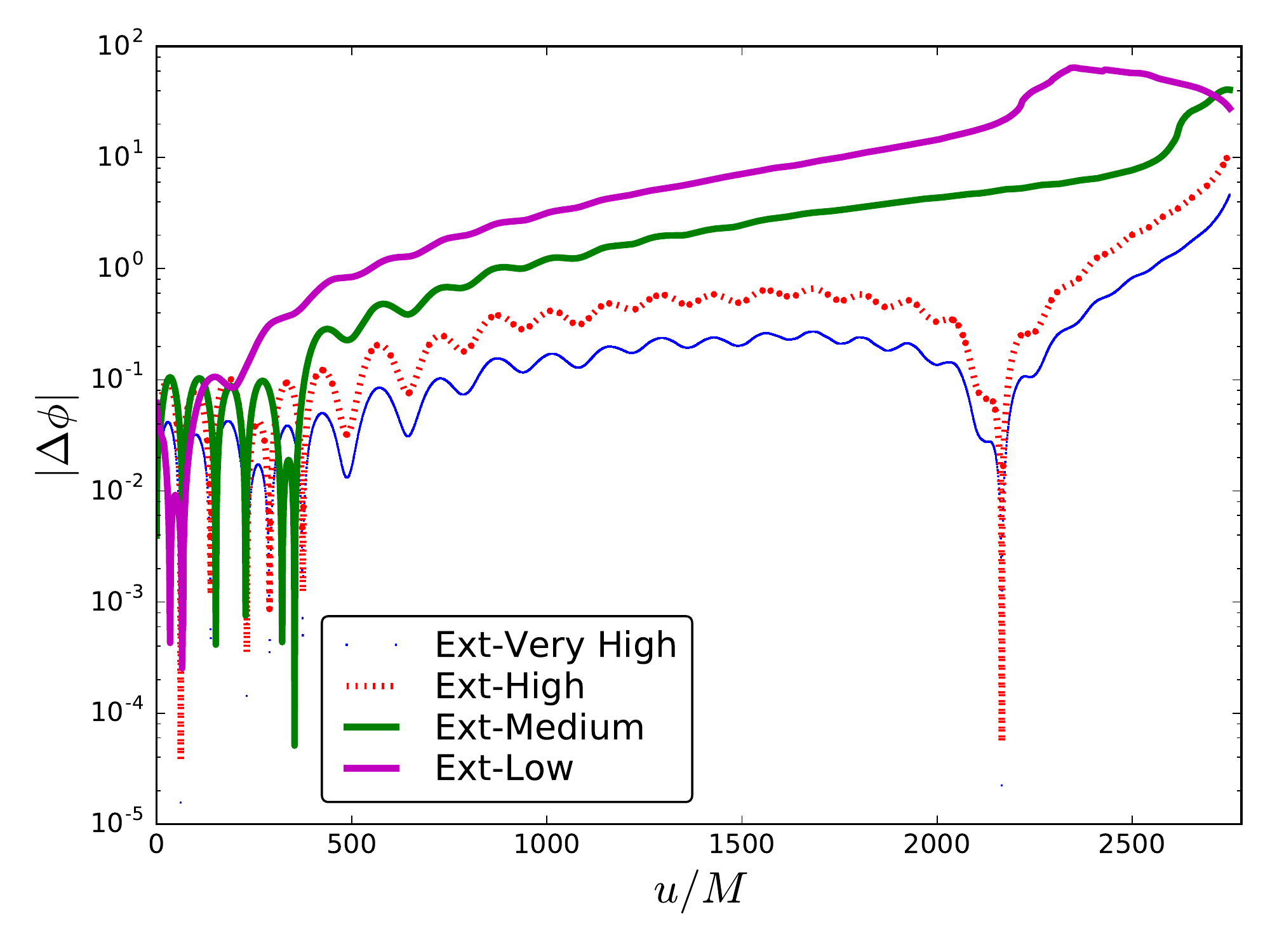}
	\caption{Phase errors relative to the Richardson extrapolated phase for the \sam SLy case. 
The differences between the extrapolated phase and the various resolutions are shown. That they decrease with resolution is another indication of convergence.
}
	\label{fig:phaseerrorextrap}
\end{figure}

Assuming a constant convergence factor of $4$,
we can calculate the Richardson extrapolation
of the GW phase. By subtracting this phase from each of phases obtained
in our simulations, we get another measure of the phase errors. 
As we display in Fig.~\ref{fig:phaseerrorextrap}, these errors for such long waveforms is of the order of 1 radian over 12 cycles, although it remain below 0.1 radians for most of the coalescence and it only increases in the last 2-3 orbits.

%%%%%%%%%%%%%%%%%%%

\begin{figure}
\includegraphics[height=2.6in]{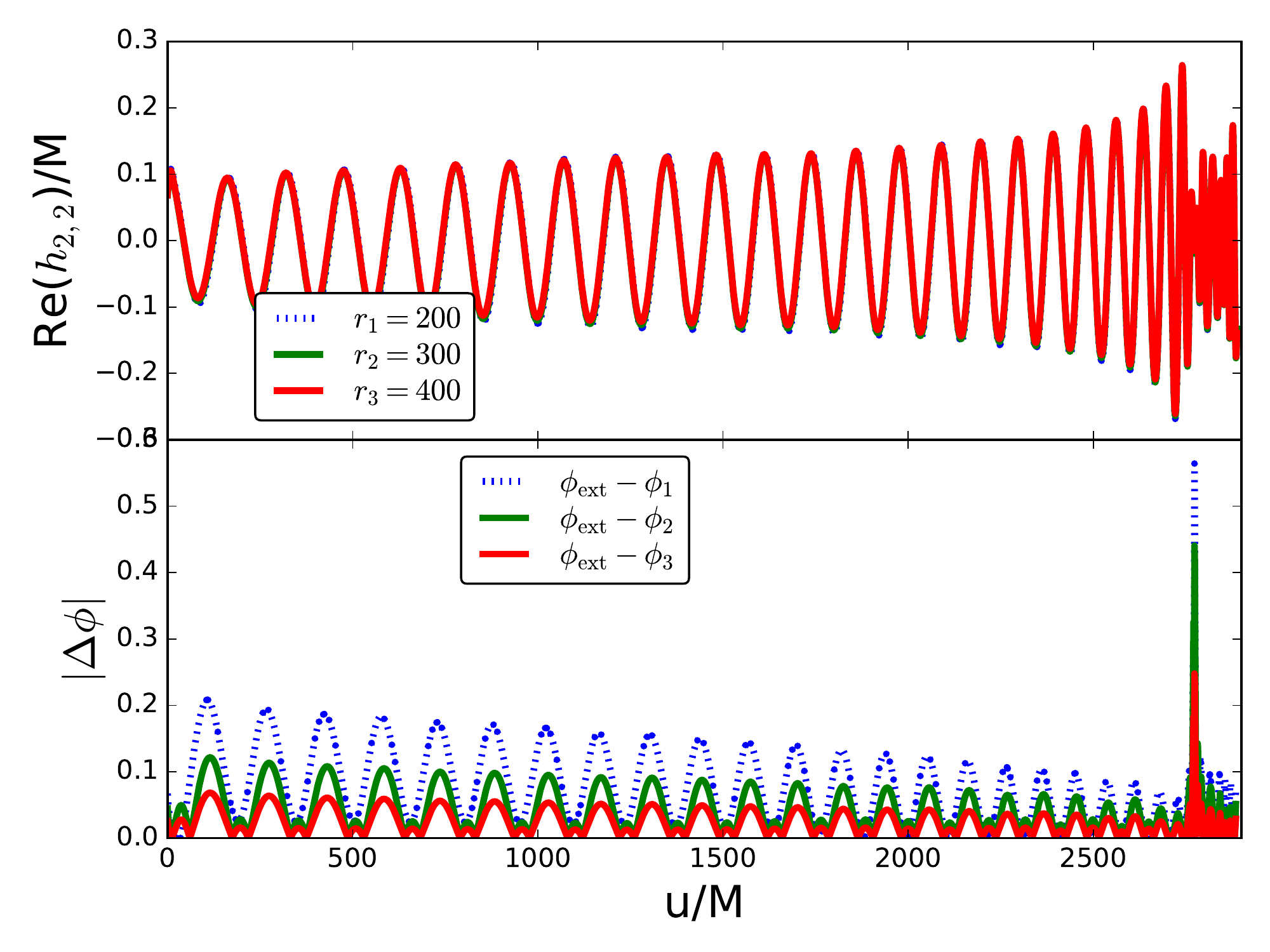}
\caption{Results from the three different extraction surfaces for the 
finest SLy run. The top panel shows the strain and the bottom panel shows
the differences between the phase obtained on each surface and the
phase obtained by extrapolating to infinite radius with the outermost
two surfaces.
}
\label{fig:surfaces}
\end{figure}

We can also analyze the various radiative quantities that are extracted on a spherical surface. 
We extract on three spherical surfaces at the different radii $r=\{350,450,600\}$ km, and plot those results for the finest 
resolution SLy run in Fig.~\ref{fig:surfaces}. That these quantities largely agree gives some indication that their extraction is correct. We can make a quantitative statement by extrapolating 
the waveforms to infinity and subtracting each of the signals obtained
on each surface. This measure indicates that the error due to the
finite size extraction radii are much smaller than the discretization
errors.

\begin{figure}
\includegraphics[height=2.6in]{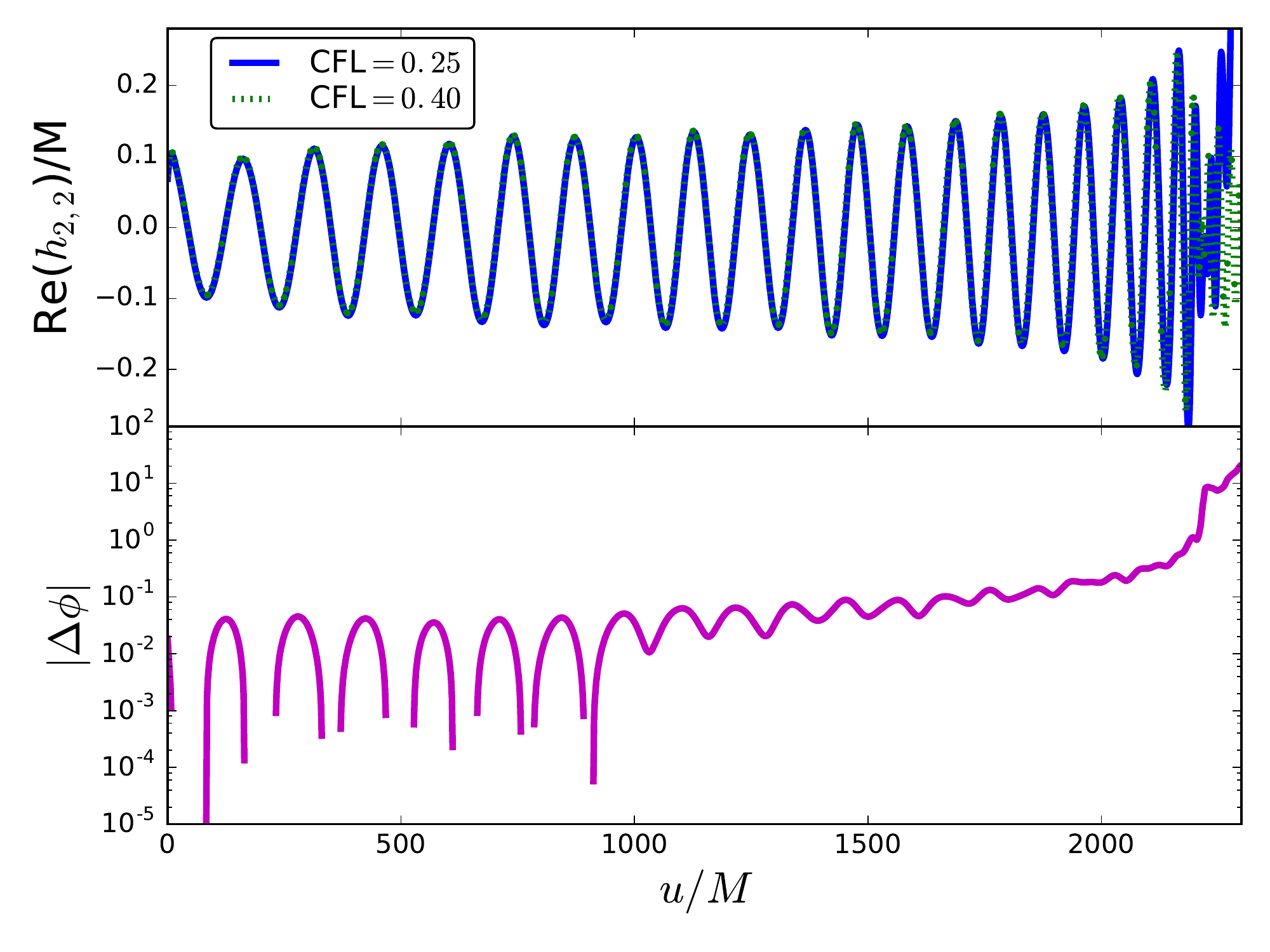}
\caption{Phase error for the \sam SLy low resolution runs with
two different CFL ratios.
That the phase difference (shown in the bottom panel) is small,
compared to Fig.~\ref{fig:phaseerrors}, indicate that the use
of a CFL factor of $0.40$ is not causing a large error in the solution.
The ability to run with a high CFL factor is itself an efficient aspect of
the \sam code.
}
\label{fig:cfl}
\end{figure}

\begin{figure}[h]
             \includegraphics[height=2.6in]{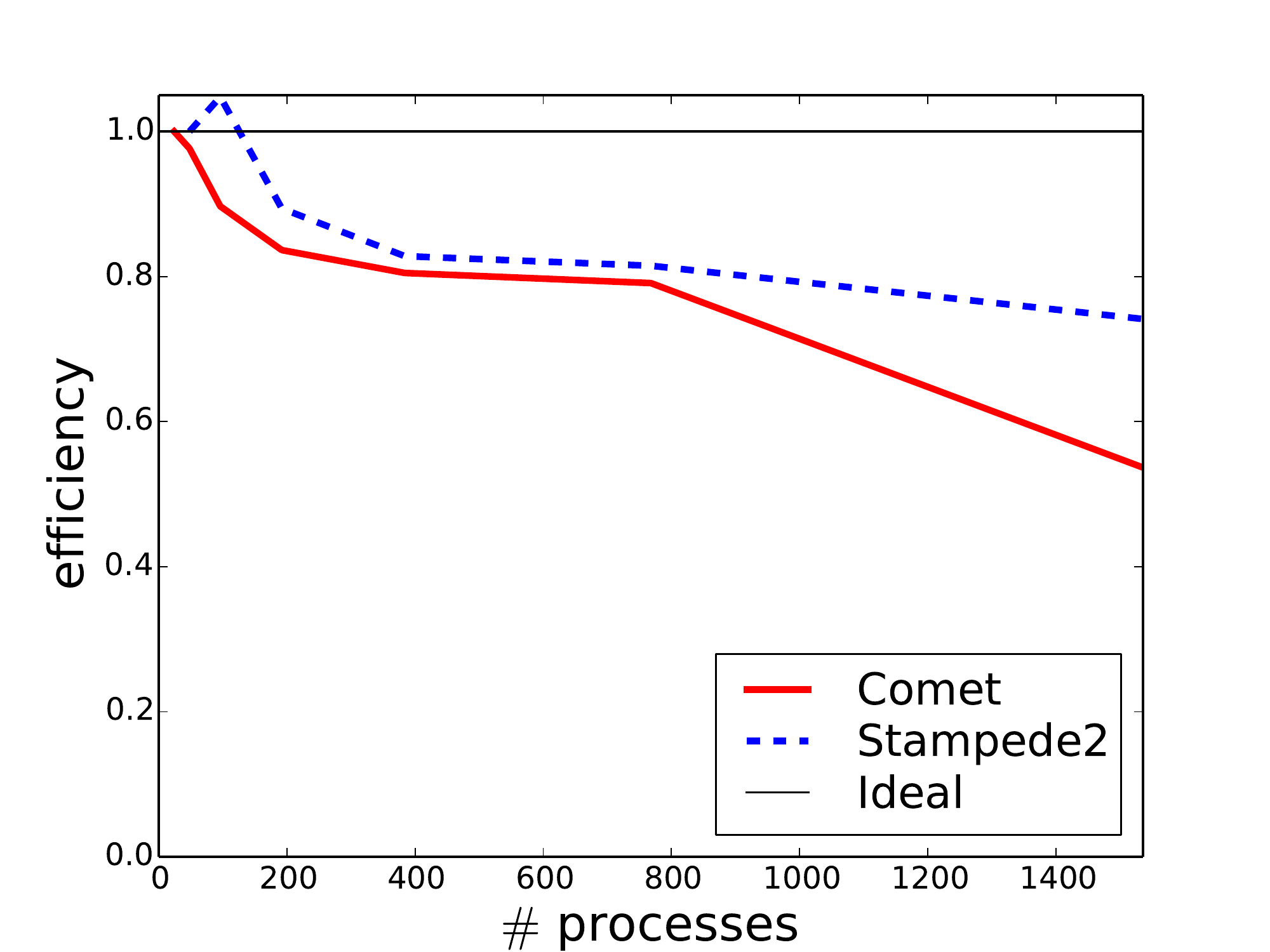}
\caption{
     Weak scaling test of the \sam code on {\tt stampede2} and
{\tt comet}. This run used fixed meshes with two levels. 
     More recent tests have indicated (see Figure 15 of ~\cite{Palenzuela:2018sly}) scaling efficiencies of 80\% extend to at least 10,000 processes. 
}
\label{fig:weak}
\end{figure}

{To examine the errors associated with our use of a large CFL factor $\lambda_c = 0.4$ (as compared to other work using similar numerical schemes in this community~\cite{Radice:2013xpa,Bernuzzi:2016pie} using $\lambda_c\approx0.25$), we adopt these two values with a low resolution run of the SLy EOS with the \sam code and compare the results in Fig.~\ref{fig:cfl}. The phase difference between the two simulations suggests a small error until merger, as compared to the phase errors between different resolutions displayed in Fig.~\ref{fig:phaseerrors}. This comparison further suggests that accurate (and efficient) simulations can be performed with the \sam code  using such a large CFL factor.}

%%%%%%%%%%%%%%%%%%%%%%%%%%%%%%%%%%%%%%%%%%%%%%%%%%%%%%%%%%%%%%%
%%%%%%%%%%%%%%%%%%%%%%%%%%%%%%%%%%%%%%%%%%%%%%%%%%%%%%%%%%%%%%%
\section{Summary}
%%%%%%%%%%%%%%%%%%%%%%%%%%%%%%%%%%%%%%%%%%%%%%%%%%%%%%%%%%%%%%%
%%%%%%%%%%%%%%%%%%%%%%%%%%%%%%%%%%%%%%%%%%%%%%%%%%%%%%%%%%%%%%%

We have presented binary neutron star simulations using two different EOS to compare two codes implemented in separate computational infrastructures. These codes solve the same evolution equations for both the Einstein and the MHD equations, and share the numerical schemes for smooth solutions. However, they implement different High-Resolution-Shock-Capturing methods: while \had uses an approach strongly inspired on finite-volume methods, 
where the fields are reconstructed at the interfaces and a flux-formula is used for calculating the fluxes at such interfaces, \sam relies on finite-difference methods, which involve a high order reconstruction of a combination of fluxes and fields (i.e., Lax-Friedrichs flux splitting). Another important difference is the treatment of the AMR boundaries, which is more efficient in \sam due to the use of a minimum bandwidth. Finally \sam is built using the public infrastructure \samraiX, which has been developed for a number years and which can reach the exascale, at least for some choice of problems. For our main purposes, that of compact binary simulations, we have achieved 
good scaling up to 10,000 processes with our high-order schemes
(see Fig. \ref{fig:weak} for weak scaling results with up to 1,500 processes with the GRMHD code
and Fig. 15 of \cite{Palenzuela:2018sly}).

The results presented here show several features. First, we have shown that the solutions of both codes are consistent and agree up to certain level. A more detailed analysis of one of these cases with \sam indicates that the solution roughly converges to fourth order. Richardson extrapolation allows us to estimate the errors in the phase, which remain below 0.2 radians for most of the coalescence for the highest resolution  case with $\Delta x = 160m$ and only increases to order unity in the last few orbits.

Another important technical result is that the speed and efficiency of the \sam code is better than \hadX, being able to achieve speeds of $\approx 20$ $M_{\odot}$/hour when running on 500-1000 processors for the highest resolution simulations of binary neutron stars presented here. This speed is roughly 4 times faster than the one reached by \hadX, which is moreover running at a resolution $\approx 15\%$ lower in our finest resolution comparison. For instance,
on comparable architectures, \sam does roughly $85$ms per month running on 192 processors 
with a minimum resolution of $\Delta x= 0.21$, while \hadX~takes approximately
$20$ms per month on 256 processors with even a slightly coarser minimum resolution of $\Delta x= 0.25$.

Of course, much still needs to be added to \samX. In particular, the ability to handle tabulated equation
of state, an approximate neutrino treatment and resistive magneto hydrodynamics, which are already available in \hadX. These physics ingredients are gradually being incorporated in \samX. Certainly, results presented here are encouraging for this enterprise.

\smallskip

\begin{acknowledgments}
We thank Will East and David Neilsen for discussions during this work.
This work was supported by the NSF under grants 
PHY-1827573 
and 
PHY-1912769.
CP acknowledges support from the Spanish Ministry of Economy and Competitiveness grant AYA2016-80289-P (AEI/FEDER, UE). 
LL was supported in part by NSERC through a Discovery Grant, and CIFAR.  
Computations were performed at XSEDE, Marenostrum and the Niagara supercomputer at the SciNet HPC Consortium. 
Computer resources at MareNostrum and the technical support provided by Barcelona Supercomputing Center were
obtained thanks to time granted through the $17^{th}$ PRACE regular call (project Tier-0 GEEFBNSM, P.I. CP).  
SciNet is funded by: the Canada Foundation for Innovation; the Government of Ontario; Ontario Research Fund - Research Excellence; and the University of Toronto.
Research at Perimeter Institute is supported by the Government of Canada and by the Province of Ontario
through the Ministry of Research, Innovation and Science.

\end{acknowledgments}

\bibliographystyle{utphys}
\bibliography{paper}

\end{document}